\documentclass[12pt,reqno]{article}
\usepackage{amsfonts}
\usepackage{amsmath}
\usepackage{amsbsy}

\usepackage{amssymb,latexsym}
\numberwithin{equation}{section}

\begin{document}
\allowdisplaybreaks[1]
\title{Dualisation of the Symmetric Space Sigma Model with Couplings}
\author{Tekin Dereli\\
Department of Physics,\\
Ko\c{c} University,\\
Rumelifeneri Yolu 34450,\\
Sar$\i$yer, \.{I}stanbul, Turkey.\\
\texttt{tdereli@ku.edu.tr},\\
\\
Nejat T. Y$\i$lmaz\\
Department of Mathematics and Computer Science,\\
\c{C}ankaya University,\\
\"{O}\u{g}retmenler Cad. No:14 06530,\\
Balgat, Ankara, Turkey.\\
\texttt{ntyilmaz@cankaya.edu.tr}} \maketitle
\begin{abstract}The first-order formulation of the $G/K$ symmetric space sigma model of
the scalar cosets of the supergravity theories is discussed when
there is coupling of ($m-1$)-form matter fields. The Lie
superalgebra which enables the dualized coset formulation is
constructed for a general scalar coset $G/K$ with matter coupling
where $G$ is a non-compact real form of a semi-simple Lie group
and $K$ is its maximal compact subgroup.

\end{abstract}

\section{Introduction}

The non-linear nature of the scalar sectors of the maximal
supergravities has been enlarged to formulate the
non-gravitational bosonic field equations as non-linear
realizations in \cite{julia1,julia2}. The coset formulation of the
scalars is improved to cover the other bosonic fields as well. The
method of \cite{julia1,julia2} includes the dualisation of the
field content and the construction of a Lie superalgebra which
generates the doubled coset element whose Cartan form would lead
to the original field equations by satisfying the Cartan-Maurer
equation. After the determination of the algebra structure it is
possible to express the first-order field equations as a twisted
self-duality condition which the dualized Cartan form satisfies.
In \cite{west1,west2,west3} a more general coset formulation of
the IIA \cite{2A1,2A2,2A3}, the IIB \cite{2B1,2B2,2B3} and the
D=11 \cite{d=11} supergravity theories is introduced to include
the gravity as well.

The scalar sectors of a wide class of supergravities, in
particular the scalar sectors of all the pure and the matter
coupled $N>2$ extended supergravities in $D=4,5,6,7,8,9$
dimensions as well as the maximally extended supergravities in
$D\leq11$ can be formulated as symmetric space sigma models. The
global symmetry groups $G$ of the scalar also the bosonic sectors
of the lower dimensional Kaluza-Klein descendant supergravities of
the D=11 supergravity (the maximal supergravities) are semi-simple
split real forms (maximally non-compact). For this reason the
scalar coset manifolds $G/K$ where $K$ is the maximal compact
subgroup of $G$ are Riemannian globally symmetric spaces
\cite{hel} and they can be parameterized by the Borel subalgebra
of $G$. In general, especially for the matter coupled
supergravities, the scalar coset manifolds $G/K$ are based on
non-split \footnote{By non-split we mean that $G$ is a non-compact
real form of a semi-simple Lie group but it is not necessarily
maximally non-compact (split).} global symmetry groups $G$. In
this case one has to use the solvable Lie algebra gauge \cite{fre}
to parameterize the Riemannian globally symmetric space scalar
coset manifold $G/K$.

In \cite{nej2} the $G/K$ symmetric space sigma model is discussed
in detail when the global symmetry group $G$ is in general, a
non-compact semi-simple real form. The dualisation and the
first-order formulation of the general non-split symmetric space
sigma model is also performed in \cite{nej2}. In this work we
consider the coupling of other fields to the scalar coset
Lagrangian of the general non-split $G/K$ symmetric space sigma
model. We will perform the complete dualisation of the fields and
the first-order formulation when there is coupling of other
($m-1$)-form matter fields to the scalar coset $G/K$. We will
construct the dualized coset element which will realize the field
equations of the scalar coset which is coupled to the ($m-1$)-form
fields. We will assume the most general non-split scalar coset
case which is discussed in \cite{nej2,ker2,ker1}. Beside the
scalar fields there will be a number of $m$-form field strengths
whose number is fixed by the dimension of the fundamental
representation of the Lie algebra $\mathbf{g}_{0}$ of $G$. As it
will be clear in the next section the dimension of the
representation and the number of the coupling fields must be the
same so that the coupling kinetic term between the scalar coset
and the matter fields in the lagrangian can be constructed within
an appropriate representation of the global symmetry group $G$
\cite{ker2,ker1}. We will follow the standard dualisation method
of \cite{julia1,julia2} by introducing auxiliary dual fields and
by assigning generators to the original and the dual fields. The
first objective of this work will be to derive the Lie
superalgebra structure which generates the doubled coset element.
The first-order formulation will then be presented as a twisted
self-duality equation \cite{julia1,julia2} by using the derived
algebra structure and by calculating explicitly the doubled field
strength. The dualisation method presented in \cite{julia1,julia2}
is the non-linear realization of the relative supergravity theory,
it is also another manifestation of the Lagrange multiplier
methods in which the dual fields correspond to the Lagrange
multipliers which are introduced to construct the Bianchi
Lagrangians. For this reason the Cartan form which is generated by
the dualized coset element, not only realizes the original
second-order field equations of the matter coupled scalar coset by
satisfying the Cartan-Maurer equation but also yields the
first-order field equations via a twisted self-duality equation
\cite{julia1,julia2,nej2}. This first-order formulation
corresponds to the construction of the dualized Lagrangian by
adding the Bianchi terms to the Lagrangian of the original fields
and consequently to the derivation of the first-order algebraic
field equations of the original fields in terms of the Lagrange
multiplier (dual) fields \cite{pope}.

 We start by discussing the Lagrangian and deriving
the field equations in Section two. In Section three we work out
the dualisation and we construct the algebraic structure which
realizes the field equations and finally we obtain the first-order
field equations.
\section{The Symmetric Space Sigma Model and the Couplings}
The scalar sectors of a wide class of supergravity theories are
formulated as $G/K$ symmetric space sigma models
\cite{julia1,julia2,ker2,ker1}. The group $G$ is the global
symmetry group of the corresponding scalar Lagrangian and it is a
non-compact real form of a semi-simple Lie group. The local
symmetry group $K$ is the maximal compact subgroup of $G$. The
coset space $G/K$ is a Riemannian globally symmetric space for all
the possible $G$-invariant Riemannian structures on $G/K$
\cite{hel}. There is a legitimate parametrization of the coset
representatives by using the solvable Lie algebra of $G$
\cite{hel,fre}. If $\mathbf{h_{k}}$ is the subalgebra of the
Cartan subalgebra $\mathbf{h}_{0}$ of $\mathbf{g}_{0}$ (the Lie
algebra of $G$) which generates the maximal R-split torus in $G$
\cite{hel,fre,nej2,ker2} let for $i=1,...,r$ $\{H_{i}\}$ be the
generators of $\mathbf{h_{k}}$ and also let $\{E_{m}\}$ be the
subset of the positive root generators of $\mathbf{g}_{0}$ such
that $m\in\Delta_{nc}^{+}$. The roots in $\Delta_{nc}^{+}$ are the
non-compact roots with respect to the Cartan involution $\theta$
which is induced by the Cartan decomposition
\begin{equation}\label{cartandecomp}
\mathbf{g}_{0}=\mathbf{k}_{0}\oplus\mathbf{u}_{0},
\end{equation}
where $\mathbf{k}_{0}$ is the Lie algebra of $K$ and
$\mathbf{u}_{0}$ is a vector subspace of $\mathbf{g}_{0}$
\cite{hel,nej2}. The positive root generators $\{E_{m}\}$ generate
a nilpotent Lie subalgebra $\mathbf{n_{k}}$ of $\mathbf{g}_{0}$
\cite{ker2}. The coset representatives of $G/K$ which are the
image points of the map from the $D$-dimensional spacetime (we
assume D$\:>2$ in order that the dualisation analysis of the next
section would be meaningful and we will take the signature of the
spacetime as ($-,+,+,...$) ) into the group $G$ can be expressed
as
\begin{equation}\label{nu}
\nu (x)=e^{\frac{1}{2}\phi ^{i}(x)H_{i}}e^{\chi ^{m}(x)E_{m}}.
\end{equation}
This is called the solvable Lie algebra parametrization
\cite{fre}. We should state that we make use of the Iwasawa
decomposition
\begin{subequations}\label{iwasawa}
\begin{align}
\mathbf{g}_{0}&=\mathbf{k}_{0}\oplus \mathbf{s}_{0}\notag\\
\notag\\
&=\mathbf{k}_{0}\oplus \mathbf{h_{k}}\oplus
\mathbf{n_{k}},\tag{\ref{iwasawa}}
\end{align}
\end{subequations}
where $\mathbf{s}_{0}$ is the solvable Lie subalgebra of
$\mathbf{g}_{0}$ which is isomorphic to $\mathbf{u}_{0}$ as a
vector space \cite{hel,nej2}. The difeomorphism from
$\mathbf{u}_{0}$ onto the Riemannian globally symmetric space
$G/K$ \cite{hel} enables the construction of the parametrization
in \eqref{nu}.

An involutive automorphism $\theta\in Aut(\mathbf{g}_{0})$ of a
semi-simple real Lie algebra $\mathbf{g}_{0}$ is called a Cartan
involution if the induced bilinear form
$B_{\theta}(X,Y)=-B(X,\theta(Y))$ where $B$ is the Killing form on
$\mathbf{g}_{0}$ is strictly positive definite $\forall X,Y\in
\mathbf{g}_{0}$. If the semi-simple complex Lie algebra
$\mathbf{g}=\mathbf{g}_{0}^{C}$ is the complexification of
$\mathbf{g}_{0}$ then the set of elements $\mathbf{t}$ of
$\mathbf{g}$ which is generated as
\begin{equation}\label{ch3315}
\mathbf{t}=\mathbf{k}_{0}+i\mathbf{u}_{0},
\end{equation}
through the complexification of $\mathbf{g}_{0}$, is a compact
real form of $\mathbf{g}$ whose conjugation will be denoted by
$\tau$. We should bear in mind that $\mathbf{g}_{0}$ has the set
equivalent images in $\mathbf{g}=\mathbf{g}_{0}^{C}$ whose
realizations in $\mathbf{g}_{0}\times \mathbf{g}_{0}$ are
isomorphic to $\mathbf{g}_{0}$. In this way $\mathbf{k}_{0}$ and
$\mathbf{u}_{0}$ can be considered as subsets of $\mathbf{g}$ and
then $\mathbf{t}$ which is a subset of $\mathbf{g}$ is also a
subset of one of the images of $\mathbf{g}_{0}$ in $\mathbf{g}$.
Thus under the realization of $\mathbf{g}$, $\,\mathbf{t}^{R}$
corresponds to a subalgebra of $\mathbf{g}_{0}$. The real
semi-simple Lie algebra $\mathbf{g}_{0}$ is also a real form of
its complexification $\mathbf{g}$ so that we may define $\sigma$
as the conjugation of $\mathbf{g}$ with respect to
$\mathbf{g}_{0}$. The map $\theta=\sigma\cdot \tau=\tau\cdot
\sigma$ is an involutive automorphism of $\mathbf{g}$. In fact
$\theta$ is a Cartan involution of $\mathbf{g}$. The
${\Bbb{R}}$-linear restriction of $\theta$ on the image of
$\mathbf{g}_{0}$ in $\mathbf{g}$ induces a Cartan involution on
$\mathbf{g}_{0}$ which we will again denote by $\theta$. After the
introduction of the Cartan involution $\theta$ we can easily
define the roots in $\Delta_{nc}^{+}$. For each element $\alpha\in
\mathbf{h}_{0}^{\ast}$ the dual space of the Cartan subalgebra $
\mathbf{h}_{0}$ of $\mathbf{g}_{0}$ we can define the element
$\alpha^{\theta}\in \mathbf{h}_{0}^{\ast}$ such that
$\alpha^{\theta}(H)=\alpha(\theta(H))$, $\,\forall \,H\in
\mathbf{h}_{0}$. If $\alpha\in \Delta$ then $\alpha^{\theta}\in
\Delta$ as well. Thus we have defined
\begin{equation}\label{delta}
\Delta_{nc}^{+}=\{\alpha\mid
\alpha\in\Delta^{+},\,\alpha\neq\alpha^{\theta}\}.
\end{equation}
The scalar Lagrangian is defined in terms of the internal metric
$\mathcal{M=}\nu ^{\#}\nu $ where we have introduced the
generalized transpose $\#$ which is over the Lie group $G$ such
that $(exp(g))^{\#}=exp(g^{\#})$ $\forall g\in \mathbf{g}_{0}$. It
is induced by the Cartan involution $\theta$ over the Lie algebra
$\mathbf{g}_{0}$ of $G$ ($g^{\#}=-\theta(g)$)
\cite{julia1,ker1,nej2}. Thus in terms of the internal metric
$\mathcal{M}$ the globally $G$-invariant and the locally
$K$-invariant scalar Lagrangian \cite{julia1,julia2} is
\begin{equation}\label{scalag}
 \mathcal{L}_{scalar}=\frac{1}{4}tr(d\mathcal{M}^{-1}\wedge
\ast d\mathcal{M}).
\end{equation}
The $G/K$ symmetric space sigma model is studied in detail in
\cite{nej2}. Thus referring to \cite{nej2} we can calculate the
Cartan form $\mathcal{G}_{0}=d\nu \nu ^{-1}$ generated by the map
\eqref{nu} as
\begin{equation}\label{g0}
\mathcal{G}_{0}=\frac{1}{2}d\phi
^{i}H_{i}+\overset{\rightharpoonup }{ \mathbf{E}^{\prime
}}\:\mathbf{\Omega }\:\overset{\rightharpoonup }{d\chi }.
\end{equation}
We have used that $[H_{i},E_{\alpha}]=\alpha_{i}E_{\alpha}$. The
row vector $\overset{\rightharpoonup }{ \mathbf{E}^{\prime }}$ has
the components $(\overset{\rightharpoonup }{
\mathbf{E}^{\prime }})_{\alpha}=e^{%
\frac{1}{2}\alpha _{i}\phi ^{i}}E_{\alpha}$. The column vector
$\overset{\rightharpoonup }{d\chi }$ is ($d\chi ^{\alpha}$). We
have also defined the matrix $\mathbf{\Omega}$ as
\begin{equation}\label{omega}
\begin{aligned}
 \mathbf{\Omega}&=\sum\limits_{n=0}^{\infty }\dfrac{\omega
^{n}}{(n+1)!}\\
\\
&=(e^{\omega}-I)\,\omega^{-1}
\end{aligned}
\end{equation}
where $\omega _{\beta }^{\gamma }=\chi ^{\alpha }\,K_{\alpha \beta
}^{\gamma }$ with the structure constants $K_{\alpha \beta
}^{\gamma }$ defined as $[E_{\alpha },E_{\beta }]=K_{\alpha \beta
}^{\gamma }\,E_{\gamma }$. Here both $\mathbf{\Omega}$ and
$\omega$ are $n\times n$ matrices where $n$ is the number of the
roots in $\Delta_{nc}^{+}$ \cite{nej2}.

We will consider the coupling of $(m-1)$-form potential fields
$\{A^{l}\}$ to the $G/K$ scalar coset where the number of the
coupling fields is determined such that they form a fundamental
representation of $\mathbf{g}_{0}$. The quadratic terms due to
this coupling which must be added to the scalar Lagrangian
\eqref{scalag} are the combinations of the internal metric
$\mathcal{M}$ and the field strengths $F^{l}=dA^{l}$
\begin{equation}\label{lagm}
\begin{aligned}
 \mathcal{L}_{m}&=-\frac{1}{2}\mathcal{M}_{kl} F^{k}\wedge\ast
 F^{l}\\
\\
&=-\frac{1}{2}F\wedge\mathcal{M} \ast F.
\end{aligned}
\end{equation}
As it is clear from above $\mathcal{M}$ and $\nu$ are in an
appropriate representation (i.e. fundamental representation of
$\mathbf{g}_{0}$) which is compatible with the number of the
coupling fields. Thus the total Lagrangian becomes
\begin{equation}\label{lag}
 \mathcal{L}=\frac{1}{4}tr( d\mathcal{M}^{-1}\wedge \ast d\mathcal{M})
 -\frac{1}{2}F\wedge\mathcal{M} \ast F.
\end{equation}
The Cartan involution $\theta$ induced by the Cartan decomposition
\eqref{cartandecomp} is an involutive automorphism of
$\mathbf{g}_{0}$ for this reason it has two eigenspaces
$\theta^{+}$, $\theta^{-}$ with eigenvalues $\pm 1$. The Cartan
involution $\theta$ induces the eigenspace decomposition of the
Lie algebra $\mathbf{g}_{0}$ as
\begin{equation}\label{eigen}
\mathbf{g}_{0}=\theta^{+}\oplus \theta^{-}.
\end{equation}
The elements of $\theta^{+}$ are called compact while the elements
of $\theta^{-}$ are called non-compact. If the subgroup of $G$
generated by the compact generators is an orthogonal group then in
the fundamental representation the generators can be chosen such
that $\mathbf{g}^{\#}=\mathbf{g}^{T}$. Therefore $\#$ coincides
with the ordinary matrix transpose and $\mathcal{M}$ becomes a
symmetric matrix in the representation we choose. We will assume
this case in our further analysis bearing in mind that for the
general case higher dimensional representations are possible in
which we can still take $\mathbf{g}^{\#}=\mathbf{g}^{T}$
\cite{ker1}.

By following the analysis of \cite{nej2,ker2,ker1} and by using
\eqref{g0} we can derive the field equations for the coupling
potentials $\{A^{k}\}$, the axions $\{\chi^{m}\}$ and the dilatons
$\{\phi^{i}\}$ of the Lagrangian \eqref{lag}. Thus the
corresponding field equations are
\begin{equation}\label{fielde}
\begin{aligned}
d(\mathcal{M}_{kl}\ast F^{l})&=0,\\
\\
 d(e^{\frac{1}{2}\gamma
_{i}\phi ^{i}}\ast U^{\gamma })&=-\frac{1}{2}\gamma
_{j}e^{\frac{1}{2}\gamma _{i}\phi
^{i}}d\phi ^{j}\wedge \ast U^{\gamma }\\
\\
&+\sum\limits_{\alpha -\beta =-\gamma }e^{\frac{1}{2} \alpha
_{i}\phi ^{i}}e^{\frac{1}{2}\beta _{i}\phi ^{i}}N_{\alpha ,-\beta
}U^{\alpha }\wedge \ast U^{\beta
},\\
\\
d(\ast d\phi ^{i})&=\frac{1}{2}\sum\limits_{\alpha\in\Delta_{nc}^{+}}^{}\alpha _{i}%
e^{\frac{1}{2}\alpha _{i}\phi ^{i}}U^{\alpha }\wedge e^{%
\frac{1}{2}\alpha _{i}\phi ^{i}}\ast U^{\alpha
}\\
\\
&+(-1)^{D+1}\frac{1}{2}((H_{i})_{nl}\nu_{m}^{n}\nu_{j}^{l})F^{j}\wedge\ast
F^{m}
\end{aligned}
\end{equation}
where $i,j=1,...,r$ and $\alpha,\beta,\gamma\in\Delta_{nc}^{+}$.
The roots in $\Delta_{nc}^{+}$ and their corresponding generators
$\{E_{m}\}$ are assumed to be enumerated. We have also defined the
vector
$U^{\alpha}=\mathbf{\Omega}_{\beta}^{\alpha}\,d\chi^{\beta}$.
Furthermore the matrices $\{(H_{i})_{nl}\}$ are the
representatives of the Cartan generators $\{H_{i}\}$ under the
representation chosen. We use the notation $[E_{\alpha },E_{\beta
}]=N_{\alpha,\beta }E_{\alpha+\beta}$. We should remark that in
the dilaton equation in \eqref{fielde} the contribution from the
coupling fields $\{A^{k}\}$ is expressed in terms of the original
fields rather than their weight expansions unlike the expressions
in \cite{ker2,ker1}. For notational convenience we raise or lower
the indices of the matrices by using an Euclidean metric.
\section{Dualisation and the First-Order Formulation}
In this section we will adopt the method of \cite{julia1,julia2}
to establish a coset formulation and to derive the first-order
field equations for the lagrangian \eqref{lag}. Basically we will
improve the analysis presented for the non-split scalar coset in
\cite{nej2} to the case when there is matter field coupling to the
non-split scalar coset. We will first define a Lie superalgebra
which will realize the doubled coset element. We assign the
generators $\{H_{i},E_{m},V_{j}\}$ to the fields
$\{\phi^{i},\chi^{m},A^{j}\}$ respectively. We assume that
$\{H_{i},E_{m}\}$ are even generators within the superalgebra
structure since the coupling fields are scalars and they have even
rank. The generators $\{V_{j}\}$ are even or odd whether the rank
of the coupling fields $\{A^{j}\}$ namely $(m-1)$ is even or odd.
The next step is to introduce the dual fields
$\{\widetilde{\phi}^{i},\widetilde{\chi}^{m},\widetilde{A}^{j}\}$
which would arise as a result of the local integration of the
field equations \eqref{fielde}. The first two are (D$-2$)-forms
and the last ones are (D$\,-m-1$)-forms. We also assign the dual
generators
$\{\widetilde{H}_{i},\widetilde{E}_{m},\widetilde{V}_{j}\}$ to
these dual fields respectively. The dual generators are even or
odd depending on D and $m$ in other words according to the rank of
the dual fields they are assigned to. We will derive the structure
of the Lie superalgebra generated by the original and the dual
generators we have introduced so that it will enable a coset
formulation for the lagrangian \eqref{lag}. Similar to the
non-linear coset structure of the scalars presented in the last
section we can define the map
\begin{equation}\label{doublenu}
\nu^{\prime}=e^{\frac{1}{2}\phi^{i}H_{i}}e^{\chi^{m}E_{m}}e^{A^{j}V_{j}}
e^{\widetilde{A}^{j}\widetilde{V}_{j
}}e^{\widetilde{\chi}^{m}\widetilde{E}_{m}}
e^{\frac{1}{2}\widetilde{\phi}^{i}\widetilde{H}_{i}},
\end{equation}
which can be considered as the parametrization of a coset via the
differential graded algebra \cite{julia2} generated by the
differential forms on the D-dimensional spacetime and the Lie
superalgebra of the original and the dual generators we propose.
We are not intending to detect the group theoretical structure of
this coset, rather we will only aim to construct the Lie
superalgebra of the original and the dual generators which
function in the parametrization \eqref{doublenu}. If one knows the
structure constants of this algebra one can calculate the Cartan
form $\mathcal{G}^{\prime}=d\nu^{\prime}\nu^{\prime-1}$ which is
induced by the map \eqref{doublenu}. Due to its definition the
Cartan form $\mathcal{G}^{\prime}$ obeys the Cartan-Maurer
equation
\begin{equation}\label{cm}
d\mathcal{G}^{\prime}-\mathcal{G}^{\prime}\wedge\mathcal{G}^{\prime}=0.
\end{equation}
By following the outline of \cite{julia1,julia2} the structure
constants of the Lie superalgebra will be chosen so that when we
calculate the Cartan form $\mathcal{G}^{\prime}$ it will lead us
to the second-order field equations \eqref{fielde} via the
identity \eqref{cm} and it will satisfy the twisted self-duality
equation $\ast\mathcal{G}^{\prime}=\mathcal{SG}^{\prime}$ where
the action of the pseudo-involution $\mathcal{S}$ \cite{julia2} on
the generators is taken as
\begin{gather}\label{s}
\mathcal{S}H_{i}=\widetilde{H}_{i}\quad,\quad\mathcal{S}E_{m}=
\widetilde{E}_{m}\quad ,\quad\mathcal{S}
\widetilde{E}_{m}=(-1)^{D}E_{m}\quad,\quad\mathcal{S}\widetilde{H}_{i}=(-1)^{D}H_{i}
,\notag\\
\notag\\
\mathcal{S}V_{j}=\widetilde{V}_{j}\quad,\quad\mathcal{S}\widetilde{V}_{j}=(-1)^{m(D-m)+1}V_{j}.
\end{gather}

We know  that the twisted self-duality equation will give us the
locally integrated first-order field equations which can be
obtained from \eqref{fielde} by extracting an overall exterior
derivative operator on both sides of the equations
\cite{julia1,julia2,nej2,pope}. This local integration produces
auxiliary fields which are the dual fields we introduce. The
dualisation method is nothing but another manifestation of the
Lagrange multiplier method while the dual fields correspond to the
Lagrange multiplier fields which are introduced to construct the
Lagrange multiplier Lagrangian terms of the Bianchi identities of
the original field strengths \cite{pope}. We may first calculate
the Cartan form $\mathcal{G}^{\prime}=d\nu^{\prime}\nu^{\prime-1}$
from the map \eqref{doublenu} in terms of the unknown structure
constants of the Lie superalgebra of the original and the dual
generators. We intend to construct an algebraic structure so that
the Cartan form satisfies the twisted self-duality equation
$\ast\mathcal{G}^{\prime}=\mathcal{SG}^{\prime}$. In a sense the
twisted self-duality equation would correspond to the equation of
motion of the dualized Lagrangian \cite{julia2}. At this stage we
will assume that the Lie superalgebra of the original and the dual
generators has a general structure in which the commutator or the
anti-commutator of two original generators gives another original
generator, an original and a dual generator leads to a dual
generator while two dual generators vanish under the algebra
product. When we calculate the structure constants of the Lie
superalgebra which generates the correct Cartan form
$\mathcal{G}^{\prime}$ which leads to the field equations
\eqref{fielde} in \eqref{cm} we will see that they obey such a
general scheme indeed. We may use the proposed twisted
self-duality property of the dualized Cartan form primarily to
write it only in terms of the original fields because as it is
clear from (3.3) the pseudo-involution sends the original
generators to the dual ones and the dual ones to the originals
with a sign factor. Thus by using the formulas
\begin{equation}\label{formul}
\begin{aligned}
de^{X}e^{-X}&=dX+\frac{1}{2!}[X,dX]+\frac{1}{3!}[X,[X,
dX]]+....,\\
\\
e^{X}Ye^{-X}&=Y+[X,Y]+\frac{1}{2!}[X,[X,Y]]+....,
\end{aligned}
\end{equation}
effectively and by applying the twisted-self duality condition
$\ast\mathcal{G}^{\prime}=\mathcal{SG}^{\prime}$, the calculation
of the Cartan form
$\mathcal{G}^{\prime}=d\nu^{\prime}\nu^{\prime-1}$ only in terms
of the original fields yields
\begin{subequations}\label{firstcartan}
\begin{align}
{\mathcal{G}}^{\prime}&=\frac{1}{2}d\phi
^{i}H_{i}+\overset{\rightharpoonup}{{\mathbf{E}}^{\prime
}}\:{\mathbf{\Omega }}\:\overset{\rightharpoonup}{d\chi
}+\overset{\rightharpoonup}{
{\mathbf{V}}}e^{{\mathbf{U}}}e^{{\mathbf{B}}}\overset{\rightharpoonup
}{{\mathbf{dA}}}\notag\\
\notag\\
&\quad +\frac{1}{2}(-1)^{D}\ast
d\phi^{i}\widetilde{H}_{i}+(-1)^{D}e^{\frac{1}{2}\alpha_{i}
\phi^{i}}{\mathbf{\Omega} }_{\beta }^{\alpha }\,\ast d\chi ^{\beta
}\widetilde{E}_{\alpha}\notag\\
\notag\\
&\quad +(-1)^{(m(D-m)+1)}\overset{\rightharpoonup
}{\widetilde{{\mathbf{V}}}}e^{{\mathbf{U}}}e^{{\mathbf{B}}}\ast\overset{\rightharpoonup
}{{\mathbf{dA}}}.\tag{\ref{firstcartan}}
\end{align}
\end{subequations}
We have defined the yet unknown structure constants as
\begin{equation}\label{structurecons2}
[H_{i},V_{n}]=\theta_{in}^{t}V_{t}\quad,\quad[E_{m},V_{j}]=\beta_{mj}^{l}V_{l}.
\end{equation}
The matrices ${\mathbf{U}}$ and ${\mathbf{B}}$ in
\eqref{firstcartan} are
\begin{equation}\label{matrice1}
({\mathbf{U}})_{v}^{n}=\frac{1}{2}\phi^{i}\theta_{iv}^{n}\quad,\quad
({\mathbf{B}})_{n}^{j}=\chi^{m}\beta_{mn}^{j}.
\end{equation}
We introduce the row vectors $\overset{\rightharpoonup
}{{\mathbf{V}}}$ and $\overset{\rightharpoonup
}{\widetilde{{\mathbf{V}}}}$ as ($V_{i}$) and
($\widetilde{V}_{j}$), respectively, the column vector
$\overset{\rightharpoonup }{{\mathbf{dA}}}$ is ($dA^{i}$). We have
also taken
\begin{equation}\label{veq}
[V_{m},V_{n}\}=0.
\end{equation}
In \eqref{firstcartan} we have made use of the results of
\cite{nej2} in the calculation of the scalar sector of the Cartan
form $\mathcal{G}^{\prime}=d\nu^{\prime}\nu^{\prime-1}$.

Now inserting the Cartan form \eqref{firstcartan} (which is
written only in terms of the original fields by primarily applying
the twisted self-duality condition) in the Cartan-Maurer identity
\eqref{cm} should result in the second-order field equations
\eqref{fielde} \cite{julia2,nej2}. This main feature of the coset
formulation enables us to derive the commutation and the
anti-commutation relations of the original generators which are
already encoded in \eqref{firstcartan} and the commutators and the
anti-commutators of the dual and the mixed (an original and a
dual) generators which arise in the calculation of \eqref{cm}
within the graded differential algebra structure of the
differential forms and the generators. Thus a straightforward
calculation of \eqref{cm} by inserting \eqref{firstcartan} and
then the comparison of the result with the second-order field
equations \eqref{fielde} gives us the desired structure constants
of the commutators and the anti-commutators. We have
\begin{gather}\label{coms}
[H_{j},E_{\alpha }]=\alpha _{j}E_{\alpha }\quad ,\quad [E_{\alpha
},E_{\beta }]=N_{\alpha ,\beta }E_{\alpha+\beta},\notag\\
\notag\\
[H_{l},V_{i}]=(H_{l})_{i}^{k}V_{k}\quad,\quad[E_{\alpha},V_{i}]=(E_{\alpha})_{i}^{j}V_{j},\notag\\
\notag\\
[H_{j},\widetilde{E}_{\alpha }]=-\alpha _{j}\widetilde{E}_{\alpha
}\quad,\quad [E_{\alpha },\widetilde{E}_{\alpha }]=\frac{1}{4}\,{\sum}_{j=1}^{r}\alpha _{j}\widetilde{H}_{j},\notag\\
\notag\\
[E_{\alpha },\widetilde{E}_{\beta }]=N_{\alpha ,-\beta }\widetilde{E}%
_{\gamma },\quad\quad\alpha -\beta =-\gamma,\;\alpha \neq
\beta,\notag\\
\notag\\
[H_{i},\widetilde{V}_{k}]=-(H_{i}^{T})_{k}^{l}\widetilde{V}_{l}\quad,\quad
[E_{\alpha},\widetilde{V}_{k}]=-(E_{\alpha}^{T})_{k}^{l}\widetilde{V}_{l},\notag\\
\notag\\
[V_{l},\widetilde{V}_{k}\}=(-1)^{D-m}\,\frac{1}{4}\,{\sum}_{i}\,(H_{i})_{lk}\,\widetilde{H}_{i},
\end{gather}
where the indices of the Cartan generators and their duals are
$i,j,l=1,...,r$ and $\alpha,\beta,\gamma\in\Delta_{nc}^{+}$. The
matrices ($(E_{\alpha})_{i}^{j}$,$(H_{l})_{i}^{j}$) above are the
representatives of the corresponding generators
($(E_{\alpha})$,$(H_{l})$). Also ($(E_{\alpha}^{T})_{i}^{j}$,
$(H_{l}^{T})_{i}^{j}$) are the matrix transpose of
($(E_{\alpha})_{i}^{j}$, $(H_{l})_{i}^{j}$). We should state once
more that the dimension of the matrices above namely the dimension
of the fundamental representation of $\mathbf{g}_{0}$ is equal to
the number of the coupling fields and their corresponding
generators since this is how we have defined and constructed the
coupling of the matter fields $A^{k}$ to the scalar coset $G/K$ in
the Lagrangian \eqref{lag}. The remaining commutators or the
anti-commutators of the original and the dual generators which are
not listed in (3.9) vanish indeed. We observe that as we have
assumed before the Lie superalgebra we have constructed in (3.9)
has the general form
\begin{gather}\label{comsorigin}
[O,\widetilde{D}\}\subset\widetilde{D}\quad,\quad [O,O\}\subset O,\notag\\
\\
[\widetilde{D},\widetilde{D}\}=0,\notag
\end{gather}
where $O$ is the set of the original and $\widetilde{D}$ is the
set of the dual generators.

Now that we have determined the structure constants of the algebra
generated by the original and the dual generators we can
explicitly calculate the Cartan form
$\mathcal{G}^{\prime}=d\nu^{\prime}\nu^{\prime-1}$in terms of both
the original and the dual fields. By using the identities in
\eqref{formul} also the structure constants given in (3.9)
effectively we have
\begin{equation}\label{expg}
\begin{aligned}
\mathcal{G}^{\prime}&=\frac{1}{2}d\phi
^{i}H_{i}+\overset{\rightharpoonup }{ \mathbf{E}^{\prime
}}\:\mathbf{\Omega }\:\overset{\rightharpoonup
}{d\chi}+\overset{\rightharpoonup
}{\widetilde{\mathbf{T}}}e^{\mathbf{\Gamma}}e^{\mathbf{\Lambda}}\overset{\rightharpoonup
}{\widetilde{\mathbf{S}}}\\
\\
&+\overset{\rightharpoonup
}{\mathbf{V}}\,\nu\,\overset{\rightharpoonup
}{\mathbf{dA}}+\overset{\rightharpoonup
}{\widetilde{\mathbf{V}}}\,(\nu^{T})^{-1}\,\overset{\rightharpoonup
}{\mathbf{d\widetilde{A}}}\\
\\
&+(-1)^{m(D-m)}\,\overset{r}{\underset{i=1}{\sum}}\,\frac{1}{4}\,(H_{i})_{kl}A^{k}\wedge
d\widetilde{A}^{l}\widetilde{H}_{i}.
\end{aligned}
\end{equation}
In addition to the definitions given in Section two we have
introduced the row vectors $\overset{\rightharpoonup
}{\mathbf{V}}$ and $\overset{\rightharpoonup
}{\widetilde{\mathbf{V}}}$ as ($V_{k}$) and ($\widetilde{V}_{l}$)
respectively. The column vectors $\overset{\rightharpoonup
}{\mathbf{dA}}$ and $\overset{\rightharpoonup
}{\mathbf{d\widetilde{A}}}$ are ($F^{k}$) and
($d\widetilde{A}^{l}$). Besides we have the row vector of the
duals of the solvable Lie algebra generators of $G$ as
$\widetilde{\mathbf{T}}_{i}=\widetilde{H}_{i}$ for $i=1,...,r$ and
$\widetilde{\mathbf{T}}_{r+\alpha}=\widetilde{E}_{\alpha}$ for
$\alpha\in\Delta_{nc}^{+}$. The column vector
$\overset{\rightharpoonup }{\widetilde{\mathbf{S}}}$ is defined as
$\widetilde{\mathbf{S}}^{i}=\frac{1}{2}d\widetilde{\phi}^{i}$ for
$i=1,...,r$ and
$\widetilde{\mathbf{S}}^{r+\alpha}=d\widetilde{\chi}^{\alpha}$ for
$\alpha\in\Delta_{nc}^{+}$. We have introduced the matrices
$\mathbf{\Gamma}$ and $\mathbf{\Lambda}$ as
\begin{equation}\label{matrice2}
\mathbf{\Gamma }%
_{n}^{k}=\frac{1}{2}\phi ^{i}\,\widetilde{g}_{in}^{k}\quad,\quad
\mathbf{\Lambda }_{n}^{k}=\chi ^{m}\widetilde{f}_{mn}^{k}.
\end{equation}
Here we have used the structure constants
$\{\widetilde{g}_{in}^{k}\}$ and $\{\widetilde{f}_{mn}^{k}\}$ from
their definitions in
\begin{equation}\label{gfstruc}
[E_{\alpha },\widetilde{T}_{m}]=\widetilde{f}_{\alpha
m}^{n}\widetilde{T}
_{n}\quad,\quad[H_{i},\widetilde{T}_{m}]=\widetilde{g}_{im}^{n}
\widetilde{T}_{n}.
\end{equation}

 They can directly be read from (3.9). If one
inserts \eqref{expg} in the Cartan-Maurer equation \eqref{cm} one
would obtain the second-order field equations and the Bianchi
identities of the original fields in terms of the original and
dual fields which are the Lagrange multipliers \cite{julia2,pope}.
One can use the twisted self-duality equation which \eqref{expg}
obeys and which gives the first-order equations to eliminate the
dual fields and then write the second-order field equations solely
in terms of the original fields namely one would reach
\eqref{fielde}. This is analogous to what we have done in the
derivation of the algebra structure. We have obtained the
second-order field equations in terms of the structure constants
of the algebra by inserting \eqref{firstcartan} in \eqref{cm} and
then we have compared the result with \eqref{fielde} to read the
structure constants. The second-order field equations in terms of
the structure constants that are mentioned above do not contain
the dual, Lagrange multiplier fields since we have used primarily
the twisted self-duality condition that relates the dual fields to
the original ones and we have written the Cartan form
$\mathcal{G}^{\prime}$ only in terms of the original fields in
\eqref{firstcartan}.

Since we have obtained the explicit form of the Cartan form
$\mathcal{G}^{\prime}$ in \eqref{expg} we can use the twisted
self-duality equation
$\ast\mathcal{G}^{\prime}=\mathcal{SG}^{\prime}$ to find the
first-order field equations of the lagrangian \eqref{lag}. The
validity of the twisted self-duality equation is justified in the
way that we have primarily assumed that $\mathcal{G}^{\prime}$
obeys it when we derived the structure constants which are chosen
such that they give the correct Cartan form $\mathcal{G}^{\prime}$
which leads to the second-order field equations \eqref{fielde} in
\eqref{cm}. Therefore directly from \eqref{expg} the twisted
self-duality equation
$\ast\mathcal{G}^{\prime}=\mathcal{SG}^{\prime}$ yields
\begin{gather}
\nu_{l}^{k}\ast
dA^{l}=(-1)^{m(D-m)+1}((\nu^{T})^{-1})_{l}^{k}d\widetilde{A}^{l},\notag\\
\notag\\
e^{\frac{1}{2}\alpha_{i}\phi^{i}}(\mathbf{\Omega})_{l}^{\alpha+r}\ast
d\chi^{l}=(-1)^{D}(e^{\mathbf{\Gamma}}
e^{\mathbf{\Lambda}})_{j}^{\alpha+r}\,\widetilde{\mathbf{S}}^{j},\notag\\
\notag\\
\frac{1}{2}\ast
d\phi^{i}=(-1)^{D}(e^{\mathbf{\Gamma}}e^{\mathbf{\Lambda}})_{j}^{i}\widetilde{\mathbf{S}}^{j}+
(-1)^{m(D-m)+D}\,\frac{1}{4}\,(H_{i})_{kl}A^{k}\wedge
d\widetilde{A}^{l}.
\end{gather}
The exterior differentiation of (3.14) gives the second-order
field equations \eqref{fielde} indeed. We should remark once more
that the roots in $\Delta_{nc}^{+}$ and the corresponding
generators $\{E_{\alpha}\}$ are enumerated. We can also express
equation (3.14) in a more compact form as
\begin{gather}
\mathcal{M}\ast \overset{\rightharpoonup
}{\mathbf{dA}}=(-1)^{m(D-m)+1}\,\overset{\rightharpoonup
}{\mathbf{d\widetilde{A}}},\notag\\
\notag\\
\ast \overset{\rightharpoonup }{\mathbf{\Psi
}}=\overset{\rightharpoonup }{\mathbf{P}}+
(-1)^{D}\,e^{\mathbf{\Gamma }%
}e^{\mathbf{\Lambda }}\overset{\rightharpoonup
}{\widetilde{\mathbf{S}}}
\end{gather}
where we define the column vector $\overset{\rightharpoonup
}{\mathbf{\Psi }}$ as
\begin{equation}\label{psi}
\begin{aligned}
\mathbf{\Psi} ^{i}&=\frac{1}{2}d\phi ^{i}\quad\quad\text{for}
\quad\quad i=1,...,r,\\
\\
\mathbf{\Psi} ^{\alpha +r}&=e^{\frac{1}{2}\alpha _{i}\phi
^{i}}\mathbf{\Omega }_{l}^{\alpha}d\chi ^{l}\quad\quad\text{for}
\quad\quad\alpha\in \Delta_{nc}^{+}.
\end{aligned}
\end{equation}
 Also the vector $\overset{\rightharpoonup }{\mathbf{P}}$ is
\begin{equation}\label{p}
\begin{aligned}
\mathbf{P}^{i}&=(-1)^{m(D-m)+D}\,\frac{1}{4}\,(H_{i})_{kl}\,A^{k}\wedge
 d\widetilde{A}^{l}\quad\quad\text{for}
\quad\quad i=1,...,r,\\
\\
\mathbf{P}^{\alpha+r}&=0\quad\quad\text{for}\quad\quad \alpha\in
\Delta_{nc}^{+}.
\end{aligned}
\end{equation}
\section{Conclusion}
After a concise discussion of the symmetric space sigma model with
its algebraic background we have defined the coupling of m-form
field strengths to the scalar Lagrangian in section two. We have
also obtained the field equations following the outline of
\cite{ker2,ker1}. In section three we have adopted the dualisation
method of \cite{julia1,julia2} to establish a coset formulation of
the theory and to explore the Lie superalgebra which leads to the
first-order equations of motion as a twisted self-duality
condition. The validity of the twisted self-duality property of
the Cartan form is implicitly justified by our construction of the
algebra since beside using the second-order field equations and
the Cartan-Maurer equation we have also assumed that the Cartan
form obeys the twisted self-duality equation in expressing it only
in terms of the original fields during the derivation of the
structure constants of the algebra. As a result we have
constructed a coset element by defining a Lie superalgebra
structure and we have shown that both the first and the
second-order field equations can be directly obtained from the
Cartan form of the coset element.

This work can be considered as an extension of the results which
are obtained in \cite{nej2}. The dualisaton of the $G/K$ symmetric
space sigma model is performed in \cite{nej2} when the global
symmetry group is a non-split semi-simple real form. Here we have
studied the dualisation of the non-split scalar coset when it is
coupled to other matter fields. We have constructed a framework in
which the dualisation analysis of \cite{nej2} is improved to
include the coupling matter fields. As a result we have obtained a
general scheme which can be effectively used in the coset
realizations of the whole set of matter coupled supergravities.

The formulation given in this work assumes a general non-split
scalar coset $G/K$ in D$\:>2$ spacetime dimensions. The coupling
potentials are assumed to be ($m-1$)-forms. As it is clear from
the construction, the results are general and they are applicable
to a wide class of supergravity theories which contain similar
couplings. In \cite{matter} the bosonic sector of the ten
dimensional simple supergravity which is coupled to $N$ Abelian
gauge multiplets is compactified on the Euclidean tori $T^{10-D}$
and the resulting theories in various dimensions have scalar
cosets with couplings based on global symmetry groups which are
non-compact real forms of some semi-simple Lie groups. Therefore
the results presented here are applicable on them.

One can improve the dualized coset formulation presented here by
including the gravity and the Chern-Simons terms as well. This
would extend the algebra structure obtained here. The group
theoretical aspects of the coset formulation and the symmetry
properties of the first-order equations which are not considered
in this work also need to be examined. One can also study the
Kac-Moody symmetry scheme \cite{west1,west2,west3} of the matter
coupled scalar cosets.


\begin{thebibliography}{99}
\bibitem{julia1}
 E. Cremmer, B. Julia, H. L\"{u} and C. N. Pope, ``\textit{Dualisation of
 dualities}", Nucl. Phys. \textbf{B523} (1998) 73, hep-th/9710119.
\bibitem{julia2}
 E. Cremmer, B. Julia, H. L\"{u} and C. N. Pope, ``\textit{Dualisation of dualities II : Twisted self-duality of doubled fields and superdualities}",
Nucl. Phys. \textbf{B535} (1998) 242, hep-th/9806106.
\bibitem{west1}
 P. C. West, ``\textit{Hidden superconformal symmetry in M theory}", JHEP
\textbf{08} (2000) 007, hep-th/0005270.
\bibitem{west2}
 P. C. West, ``\textit{E(11) and M theory}", Class. Quant. Grav. \textbf{18} (2001) 4443, hep-th/0104081.
\bibitem{west3}
 I. Schnakenburg and P. C. West, ``\textit{Kac-Moody symmetries of IIB
 supergravity}", Phys. Lett. \textbf{B517} (2001) 421, hep-th/0107181.
\bibitem{2A1}
 I. C. G. Campbell and P. C. West, ``\textit{N=2,D=10 nonchiral supergravity and its
spontaneous compactification}", Nucl. Phys. \textbf{B243} (1984)
112.
\bibitem{2A2}
 M. Huq and M. A. Namazie, ``\textit{Kaluza-Klein supergravity in ten dimensions}", Class. Quant.
 Grav. \textbf{2} (1985) 293.
\bibitem{2A3}
 F. Giani and M. Pernici, ``\textit{N=2 supergravity in ten dimensions}", Phys. Rev.
\textbf{D30} (1984) 325.
\bibitem{2B1}
 J. H. Schwarz and P. C. West, ``\textit{Symmetries and transformations of chiral N=2 D=10
 supergravity}", Phys. Lett.
\textbf{B126} (1983) 301.
\bibitem{2B2}
 P. S. Howe and P. C. West, ``\textit{The complete N=2 D=10 supergravity}", Nucl. Phys.
\textbf{B238} (1984) 181.
\bibitem{2B3}
 J. H. Schwarz, ``\textit{Covariant field equations of chiral N=2 D=10 supergravity}", Nucl. Phys.
\textbf{B226} (1983) 269.
\bibitem{d=11}
E. Cremmer, B. Julia and J. Scherk, ``\textit{Supergravity theory
in eleven-dimensions}", Phys. Lett. \textbf{B76} (1978) 409.
\bibitem{hel}
 S. Helgason ``\textbf{Differential Geometry, Lie Groups and Symmetric
 Spaces}", (Graduate Studies in Mathematics 34, American Mathematical Society
 Providence R. I. 2001).
\bibitem{fre}
L. Andrianopoli, R. D'Auria, S. Ferrara, P. Fr\'{e} and M.
Trigiante, ``\textit{RR scalars, U-duality and solvable Lie
algebras}", Nucl. Phys. \textbf{B496} (1997) 617, hep-th/9611014.
\bibitem{nej2}
 N. T. Y$\i$lmaz, ``\textit{The non-split scalar coset in supergravity theories}", Nucl.
 Phys. \textbf{B675} (2003) 122, hep-th/0407006.
\bibitem{ker2}
 A. Keurentjes, ``\textit{The group theory of oxidation II : Cosets of non-split
 groups}", Nucl. Phys. \textbf{B658} (2003) 348, hep-th/0212024.
\bibitem{ker1}
 A. Keurentjes, ``\textit{The group theory of oxidation}", Nucl. Phys.
\textbf{B658} (2003) 303, hep-th/0210178.
\bibitem{pope}
 C. N. Pope, ``\textbf{Lecture Notes on Kaluza-Klein Theory}", (unpublished).
\bibitem{matter}
H. L\"{u}, C. N. Pope and K. S. Stelle,
``\textit{M-theory/Heterotic duality: A Kaluza-Klein
perspective}", Nucl. Phys. \textbf{B548} (1999) 87,
hep-th/9810159.












\end{thebibliography}
\end{document}